\documentclass[prl,showpacs,twocolumn,aps]{revtex4}
\usepackage{graphicx}
\def\be{\begin{equation}}
\def\ee{\end{equation}}
\def\bea{\begin{eqnarray}}
\def\eea{\end{eqnarray}}

\def\Bi2212{Bi$_{2}$Sr$_{2}$CaCu$_{2}$O$_{8+\delta}$}
\def\H{{\cal H}}
\def\D{{\cal D}}

\def\etal{{\it et al.} }

\begin{document}
\title{An effective Hamiltonian for phase fluctuations on a lattice:
an extended XY model}
\author{Wonkee Kim and J. P. Carbotte}
\affiliation{Department of Physics and Astronomy,
McMaster University, Hamilton,
Ontario, Canada, L8S~4M1}
\begin{abstract}
We derive an effective Hamiltonian for phase fluctuations in an $s$-wave
superconductor starting from the attractive Hubbard model on a square
lattice.
In contrast to the common assumption, 
we find that the effective Hamiltonian is not
the usual XY model but is of an extended XY type. This extended feature
is robust and leads to essential corrections in understanding phase 
fluctuations on a lattice.
The effective coupling
in the Hamiltonian varies significantly with temperature. 

\end{abstract}
\maketitle

The physics underlying the pseudogap phenomena in the cuprates\cite{timusk} 
remains unclear and is a topic of intense debate.
One possibility is that
the pseudogap is a precursor of
the superconducting gap. This picture
finds support in
ARPES experiments\cite{ding}, which show that the pseudogap seems to evolve
into the superconducting gap as the temperature $(T)$
is lowered through the superconducting transition temperature $(T_{c})$.
In this picture, the Cooper pairs are preformed above $T_{c}$
but without
phase coherence\cite{emery}, and
the superfluid stiffness is destroyed by the phase fluctuations.
A different precursor scenario 
based on an extension of the BCS theory which induces finite momentum pairs
has also had
some success\cite{levin}. 
Since the superconductivity is believed to resides in the
CuO$_{2}$ planes 
and interlayer
coupling is weak in the cuprates, it is conceivable that the high-$T_{c}$
superconductors reflect importantly their two-dimensional nature. 
Since 
no spontaneous breaking of continuous symmetry is allowed
in 2D\cite{mermin},
a reasonable expectation is that the superconducting transition,
at least, in the underdoped cuprates is not of the BCS kind
but is rather of the 
Berezinskii-Kosterlitz-Thouless (BKT) type\cite{kt}. 

It is well-known that the effective Hamiltonian for phase fluctuations in
an $s$-wave superconductor is equivalent to $H_{XY}=J_{XY}
\int d{\bf r}(\nabla\theta)^{2}$, where $\theta$ describes phase fluctuations.
On a lattice, one has to either discretize $H_{XY}$ or derive the
effective Hamiltonian from first principle. A common recipe to 
discretize $H_{XY}$ is the assumption that the continuum 
limit still holds and, therefore, the lattice version of
the XY Hamiltonian
$(\H_{XY})
$ has been
uncritically accepted to describe phase fluctuations. 
In this paper, we derive
an effective Hamiltonian for phase fluctuations in an $s$-wave superconductor
from first principle. We find that the effective Hamiltonian
is not of $\H_{XY}$ type but rather it is an extended XY; however, its
critical behavior is still BKT-like. In contrast to the usual phenomenological 
assumption,
the effective coupling constant in the Hamiltonian now
depends significantly on temperature.

We begin with
a 2D attractive Hubbard model 
on a square lattice as the simplest case conceivable
\bea
\H=&&-t\sum_{<ij>\sigma}C^{+}_{i\sigma}C_{j\sigma}
-t'\sum_{\ll ij\gg\sigma}C^{+}_{i\sigma}C_{j\sigma}
\nonumber\\
&&-\mu\sum_{i\sigma}n_{i\sigma}
-U\sum_{i}n_{i\uparrow}n_{i\downarrow}\;,
\eea
where $C_{i\sigma}$ is a fermion field with spin $\sigma$,
$\mu$ is a chemical potential, $U(>0)$ is the pairing
interaction, 
$t$ and $t'$ describe, respectively, the nearest neighbor (n.n) and 
the next nearest neighbor (n.n.n)
hopping. The symbol $<ij>$ means a sum over the n.n pairs and
$\ll ij\gg$ indicates a sum over the n.n.n pairs.
We will set the lattice constant $a=1$ and also use units such that
$\hbar=k_{B}=1$. 
The partition function $(Z)$, base on the Hamiltonian $\H$ and
written in the language of path integral, is given by
$
Z=\int\D C^{+}\D C \exp[-S]
$,
where the action $S=\int d\tau \{\sum_{i\sigma}C^{+}_{i\sigma}
\partial_{\tau}C_{i\sigma}+\H\}$. The range for the integral
over the imaginary time
$\tau$ in the action is from $0$ to $1/T$. 
We first consider only n.n hopping $(t)$ with $t'=0$
and derive an effective local
Hamiltonian for the  phase of the order parameter.
At finite temperature 
we obtain, in this case, an extended XY model Hamiltonian 
which includes not only
the n.n spin-spin interaction but also the second and 
third neighbor interaction.
Inclusion of the n.n.n hopping makes this extended feature of the
effective Hamiltonian even more robust in the sense that this property
is now manifested even at zero temperature.

Introducing the Hubbard-Stratonovich transformation\cite{negele}
with an auxiliary field $\phi_{j}=\Delta_{j}e^{i\theta_{j}}$,
and making a gauge transformation 
$\Psi_{j}=\exp[i{\hat\tau}_{3}\theta_{j}/2]\chi_{i}$,
where $\chi_{i}$ is a spinor for neutral fermions and
${\hat\tau}_{i}$ are Pauli matrices, after integrating
out the fermion fields, we obtain $Z=\int\D\phi^{*}\D\phi \exp[-S_{eff}]$,
where an effective action is
\be
S_{eff}=\int d\tau\sum_{i}{1\over U}|\phi_{i}|^{2}-\mbox{Tr}\ln[G^{-1}]\;.
\ee
Here, $\mbox{Tr}$ means the trace over the functional space, depending on the 
representation, and spin space. We will use a symbol $\mbox{tr}$
for a trace over spin space only.
In the real (lattice) space, the Green function can be represented as
$
G^{-1}(i,j)=G^{-1}_{0}(i,j)-\Sigma(i,j)\;,
$
with
\be
G^{-1}_{0}(i,j)=(-{\hat\tau}_{0}\partial_{\tau}+{\hat\tau}_{3}\mu
+{\hat\tau}_{2}\Delta)\delta_{i,j}+
{\hat\tau}_{3}t\sum_{\bf\delta}\delta_{j,i+{\bf\delta}}
\ee
and the self energy has the form
\be
\Sigma(i,j)={i\over2}{\hat\tau}_{3}(\partial_{\tau}\theta_{j})\delta_{i,j}
+{\hat\tau}_{0}\Sigma^{(0)}(i,j)+{\hat\tau}_{3}\Sigma^{(3)}(i,j)\;,
\ee
where $\Sigma^{(0)}(i,j)=it\sum_{\bf\delta}\delta_{j,i+{\bf\delta}}
\sin(\theta_{i,j}/2)$ and $\Sigma^{(3)}(i,j)=
t\sum_{\bf\delta}\delta_{j,i+{\bf\delta}}[1-\cos(\theta_{i,j}/2)]$.
Here 
$\theta_{i,j}=\theta_{i}-\theta_{j}$ is a phase difference between
sites $i$ and $j$, and ${\bf\delta}=\pm{\hat x}$ and $\pm{\hat y}$
on a square lattice.
In the expression for $G^{-1}$
we consider only 
fluctuations of the phase and assume no fluctuation in the magnitude
of the gap.
Magnitude fluctuation will also change the self energy $\Sigma$. 
However, we ignore
this and concentrate on the effects of phase fluctuations.
For simplicity we
consider the static case $(\partial_{\tau}\theta_{i}=0)$.
A generalization to the time-dependent case is straightforward.

The effective action $S_{eff}$ can be separated into a mean-field
part $(S^{(0)})$ and a phase fluctuation part $(S_{\theta})$, namely,
$
S_{eff}=S^{(0)}+\mbox{Tr}\sum_{n=1}^{\infty}\left[G_{0}\Sigma\right]^{n}
$,
where $S^{(0)}=\int d\tau\sum_{i}{1\over U}
|\Delta|^{2}-\mbox{Tr}\ln[G^{-1}_{0}]$.
It can be shown\cite{loktev} that in the saddle-point approximation
$\partial S^{(0)}/\partial\Delta=0$ reduces to the BCS mean-field gap
equation:
$
\Delta=\sum_{\bf k}{\Delta\over E_{\bf k}}
\tanh\left({E_{\bf k}\over2T}\right)
$,
and $\partial S^{(0)}/\partial\mu$ gives an equation for the filling
factor:
$
n=1-\sum_{\bf k}{\xi_{\bf k}\over E_{\bf k}}
\tanh\left({E_{\bf k}\over2T}\right)\;,
$
where $E_{\bf k}=\sqrt{\xi^{2}_{\bf k}+\Delta^{2}}$ with
$\xi_{\bf k}=-2t\left[\cos(k_{x})+\cos(k_{y})\right]-\mu$.
When we 
include the n.n.n hopping, $\xi_{\bf k}$ has an additional term
$-4t'\cos(k_{x})\cos(k_{y})$.
In the calculation of 
the effective phase-only action $S_{\theta}$, 
we have assumed that the phase fluctuations 
are small so that we retain only up to the square of the phase
difference between two sites in expansion of $S_{\theta}$.
This means that it is sufficient to consider
only the first and the second trace for $S_{\theta}$; in other words,
$S_{\theta}\simeq S^{(1)}+S^{(2)}$, where
$
S^{(1)}=\mbox{Tr}\left[G_{0}\Sigma\right]
$,
and
$
S^{(2)}={1\over2}\mbox{Tr}\left[G_{0}\Sigma G_{0}\Sigma\right]\;.
$

For simplicity, we will use four vector notation: $K=({\bf k},\omega_{n})$,
$\sum_{K}=T\sum_{\omega_{n}}\sum_{\bf k}$, where $\omega_{n}$ is a 
fermionic Matsubara frequency, and $\int dX_{i}=\int d\tau\sum_{i}$.
It can be shown that
$\langle K|G|K'\rangle=\delta(K-K')G(K')$ and
$\langle K'|\Sigma|K\rangle={\hat\tau}_{0}{\tilde\Sigma}^{(0)}(K',K)+
{\hat\tau}_{3}{\tilde\Sigma}^{(3)}(K',K)$, where
$
{\tilde\Sigma}^{(0)}(K',K)=
it\int dX_{i}e^{-i(K'-K)X_{i}}
\sum_{\bf \delta}e^{i{\bf k}\cdot{\bf \delta}}A_{\delta}(X_{i})
$,
and
$
{\tilde\Sigma}^{(3)}(K',K)=
t\int dX_{i}e^{-i(K'-K)X_{i}}
\sum_{\bf \delta}e^{i{\bf k}\cdot{\bf \delta}}
B_{\delta}(X_{i})
$
with $A_{\delta}(X_{i})=\sin(\theta_{i,i+\delta}/2)$ and
$B_{\delta}(X_{i})=1-\cos(\theta_{i,i+\delta}/2)$.
Since $S^{(1)}=\mbox{tr}\sum_{K,K'}\langle K|G_{0}|K'\rangle
\langle K'|\Sigma|K\rangle$, we obtain
\be
S^{(1)}=t\sum_{K}\Gamma^{(3)}_{K}
\cos(k_{x})
\int d\tau\sum_{<ij>}\left[1-\cos\left({\theta_{i,j}/2}
\right)\right],
\ee
where 
$\Gamma^{(3)}_{(K)}=\mbox{tr}
\left[G(K){\hat\tau}_{3}e^{i\eta\omega_{n}{\hat\tau}_{3}}\right]$ with
$\eta\rightarrow0^{+}$ being a convergence factor\cite{abrikosov},
which originates
from $G(0,\tau-\tau^{+})$. Note that $\Sigma^{(0)}$ makes no contribution
to $S^{(1)}$. Since $S^{(1)}$ include the phase difference only between
the n.n sites and the first non-trivial term is $\theta^{2}_{i,j}$,
we obtain the usual XY model action; however, $\theta^{2}_{i,j}$ terms
will also appear in $S^{(2)}$. Moreover, the phase difference between
the n.n.n sites is also obtained from the second trace. As we will see later,
if we include the n.n.n hopping, even in $S^{(1)}$ one has the 
phase difference between the n.n.n sites.

Similarly, 
\begin{small}
$S^{(2)}={1\over2}\mbox{tr}\sum_{K,K} G(K){\tilde\Sigma}(K,K')
G(K'){\tilde\Sigma}(K',K)$, 
\end{small}
where
${\tilde\Sigma}={\hat\tau}_{0}{\tilde\Sigma}^{(0)}
+{\hat\tau}_{3}{\tilde\Sigma}^{(3)}$.
After some manipulations, one arrives at
\bea
S^{(2)}=&&{1\over2}\sum_{K,K'}
\Lambda^{(00)}_{K,K'}
\tilde{\Sigma}^{(0)}(K,K')
\tilde{\Sigma}^{(0)}(K',K)
\nonumber\\
+&&{1\over2}\sum_{K,K'}\Lambda^{(33)}_{K,K'}
\tilde{\Sigma}^{(3)}(K,K')
\tilde{\Sigma}^{(3)}(K',K)\;,
\eea
where $\Lambda^{(ii)}_{K,K'}=\mbox{tr}\left[
G(K){\hat\tau}_{i}G(K'){\hat\tau}_{i}\right]$.
The mixed terms do not contribute to $S^{(2)}$
in our consideration for a local effective action.
Since in the momentum space 
$A_{\delta}(X_{i})=\sum_{Q}{\tilde A}_{\delta}(Q)\exp[iQX_{i}]$, where
$Q$ is a four vector $({\bf q}, \Omega_{m})$ with a bosonic Matsubara
frequency $\Omega_{m}$, the first term of $S^{(2)}$ is
$-{t^{2}\over2}\sum_{\delta,\delta'}\sum_{Q}{\tilde A}_{\delta}(Q)
{\tilde A}_{\delta'}(-Q)e^{-i{\bf q}\cdot\delta}\sum_{K}
e^{i{\bf k}\cdot(\delta+\delta')}
\Lambda^{(00)}_{K,K-Q}$.
A similar expression can be obtained for the second term.
Since, as we mentioned earlier, here
we are concerned only with the local effective
action for the phase and wish to see if the equivalence between
the phase-only Hamiltonian and the usual
$H_{XY}$ found for the continuum limit also holds on a lattice,
we expand $S^{(2)}$ about $Q=0$ and keep the
leading order. This means that we neglect the so-called Landau terms
associated with damping effects, which have been shown\cite{aitchison} 
not to be important 
in a case of an
$s$-wave superconductor in the continuum limit for $T\lesssim 0.6T_{MF}$,
where $T_{MF}$ is a temperature at which $\Delta(T_{MF})=0$. 
In the $s$-wave case, 
the importance of the damping effects appears when there is a large
suppression of $\Delta(T)$ so that
$\Delta(T)$ is comparable to $T$, and 
the fraction of thermally excited quasiparticles is not negligible.
However, it has been pointed out
that for a $d$-wave superconductor the Landau terms have strong effects
even at the low temperature
because of the nodal structure of the order parameter\cite{sharapov}.
It can be shown that
the first term in $S^{(2)}$ is
${1\over2}t^{2}\sum_{\delta,\delta'}\sum_{K}\Lambda^{(00)}_{K,K}
e^{i{\bf k}\cdot(\delta-\delta')}\int dX_{i}
A_{\delta}(X_{i})A_{\delta'}(X_{i})$, and
the second term becomes
${1\over2}t^{2}\sum_{\delta,\delta'}\sum_{K}\mbox{tr}
\Lambda^{(33)}_{K,K}
e^{i{\bf k}\cdot(\delta-\delta')}\int dX_{i}
B_{\delta}(X_{i})B_{\delta'}(X_{i})$.
In order to
compare the effective Hamiltonian with the lattice version $\H_{XY}$, 
we need to expand
$S^{(2)}$ in terms of the phase differences while keeping terms up to
$\theta^{2}_{i,j}$. Since the first non-trivial term for the second 
term is $\theta^{4}_{i,j}$,
we will ignore it in the effective Hamiltonian.

Using symmetry properties of $G(K)$ such as remaining the same with respect to
the exchange of $k_{x}$ and $k_{y}$, we obtain the effective 
local Hamiltonian: $\H_{\theta}=\langle\theta|{\hat M}|\theta\rangle$,
where $\langle\theta|=(\theta_{i,i+{\hat x}},\theta_{i,i-{\hat x}},
\theta_{i,i+{\hat y}},\theta_{i,i-{\hat y}})$ and 
\begin{center}
${\hat M}=\left(\begin{array}{cccc}
         \alpha&\beta&\gamma&\gamma\\
         \beta&\alpha&\gamma&\gamma\\
         \gamma&\gamma&\alpha&\beta\\
         \gamma&\gamma&\beta&\alpha
\end{array}\right)$
\end{center}
with components as follow:
$\alpha={1\over8}t\sum_{K}\Gamma^{(3)}_{K}
\cos(k_{x})+{1\over8}t^{2}\sum_{K}\Lambda^{(00)}_{K,K}$,
$\beta={1\over8}t^{2}\sum_{K}\Lambda^{(00)}_{K,K}\cos(2k_{x})$, and
$\gamma={1\over8}t^{2}\sum_{K}
\Lambda^{(00)}_{K,K}\cos(k_{x})\cos(k_{y})$.
As one can see, the $4\times4$ matrix ${\hat M}$ is not diagonal. 
This means
that the effective Hamiltonian $\H_{\theta}$ is not equivalent to
the usual $\H_{XY}$, which would be pure-diagonal; namely, 
we have not only terms like $\theta^{2}_{i,i+{\hat x}}$ and
$\theta^{2}_{i,i-{\hat x}}$ but also terms of the form
$\theta_{i,i+{\hat x}}\theta_{i,i-{\hat x}}$ in $\H_{\theta}$.
We will next investigate effects of these off-diagonal
terms. 

It is worthwhile making sure our new result for $\H_{\theta}$ reduces
to the well known XY-type Hamiltonian\cite{loktev} in the continuum limit.
For this purpose, we need to recover the explicit value
of the lattice constant $a$ 
in the expression
for $\H_{\theta}$ in order to track orders of $a$ in the limit $a\rightarrow0$
and the phase difference becomes a derivative of the phase $(\nabla\theta)$
while $ta^{2}\rightarrow{1\over 2m}$, where
$m$ is an effective mass of an electron. It is straightforward to show
that, in the continuum limit with $\xi_{\bf k}={k^{2}\over 2m}-\mu$,
$\H_{\theta}\rightarrow H^{(1)}_{\theta}+H^{(2)}_{\theta}$, where
$H^{(1)}_{\theta}={1\over 8m}\sum_{K}\Gamma^{(3)}_{K}
\int d{\bf r}\left(\nabla\theta\right)^{2}$, and
$H^{(2)}_{\theta}={1\over 16m^{2}}\sum_{K}\Lambda^{(00)}_{K,K}k^{2}
\int d{\bf r}\left(\nabla\theta\right)^{2}$. 
Consequently, the effective Hamiltonian $\H_{\theta}$ we derive 
does reduce to
$H_{XY}$.

Let us now consider effects of the off-diagonal terms in $\H_{\theta}$.
At $T=0$, $\H_{\theta}$ again becomes equivalent to the XY-type Hamiltonian
because the off-diagonal terms of ${\hat M}$, $\beta$ and $\gamma$ vanish.
However, at finite temperature they are nonzero so that
$\H_{\theta}$ is no longer of the usual XY type.
Instead 
\bea
\H_{\theta}=&&\alpha\sum_{<ij>}\theta^{2}_{i,j}
+2\beta\sum_{i}\left(
\theta_{i,i+{\hat x}}\theta_{i,i-{\hat x}}+
\theta_{i,i+{\hat y}}\theta_{i,i-{\hat y}}\right)
\nonumber\\
+&&2\gamma\sum_{i}
\left(\theta_{i,i+{\hat x}}+\theta_{i,i-{\hat x}}\right)
\left(\theta_{i,i+{\hat y}}+\theta_{i,i-{\hat y}}\right)\;.
\eea
Since the phase fluctuation between two sites is small by
assumption,
one can make the approximation; $\theta_{i,i+{\hat x}}\simeq
\sin(\theta_{i,i+{\hat x}})$. Introducing a 2D classical spin 
${\bf S}_{i}=(\cos(\theta_{i}),\sin(\theta_{i}))$ 
at site $i$, it can be shown, within the approximation
we have made, that,
for example,
$\theta_{i,i+{\hat x}}\theta_{i,i-{\hat x}}\simeq
({\bf S}_{i}\times{\bf S}_{i+{\hat x}})\cdot
({\bf S}_{i}\times{\bf S}_{i-{\hat x}})$. Apply a vector
identity to this expression, we then obtain
$({\bf S}_{i}\times{\bf S}_{i+{\hat x}})\cdot
({\bf S}_{i}\times{\bf S}_{i-{\hat x}})={\bf S}_{i+{\hat x}}\cdot
{\bf S}_{i-{\hat x}}-({\bf S}_{i}\cdot{\bf S}_{i+{\hat x}})
({\bf S}_{i}\cdot{\bf S}_{i-{\hat x}})$, where we used
the fact that ${\bf S}_{i}\cdot{\bf S}_{i}=1$. As one can see, a spin
at the site $i+{\hat x}$ couples to a spin at $i-{\hat x}$, which
is a n.n.n spin-spin interaction. Note that the only assumption
we have made is that terms higher than $\theta^{2}_{i,j}$ for
a given $(i,j)$ are negligible.
This does not mean, however, that only the n.n phase differences are important. 
Following the procedure we briefly described, one can show that
the second term of $\H_{\theta}$, which is proportional to $\beta$, becomes
$\beta\sum_{<ij>}\theta^{2}_{i,j}+\beta\sum_{<ij>_{3}}{\bf S}_{i}\cdot
{\bf S}_{j}$ within a constant, 
where the symbol $<ij>_{3}$ indicates a sum over the next-next
nearest pairs.
Similarly, the third term of $\H_{\theta}$ proportional to $\gamma$ turns
out to be
$2\gamma\sum_{<ij>}\theta^{2}_{i,j}
+2\gamma\sum_{\ll ij\gg}{\bf S}_{i}\cdot{\bf S}_{j}$. Since
$\theta^{2}_{i,j}$ can also be represented in term of
${\bf S}_{i}\cdot{\bf S}_{j}$, the effective
Hamiltonian $\H_{\theta}$ can be written as
\begin{small}
\be
\H_{\theta}=-J_{1}\sum_{<ij>}{\bf S}_{i}\cdot{\bf S}_{j}
+J_{2}\sum_{\ll ij\gg}{\bf S}_{i}\cdot{\bf S}_{j}
+J_{3}\sum_{<ij>_{3}}{\bf S}_{i}
\cdot{\bf S}_{j}\;,
\ee
\end{small}
where $J_{1}=2(\alpha+\beta+2\gamma)$, $J_{2}=2\gamma$, and
$J_{3}=\beta$.
It is clear that this Hamiltonian is not of the usual XY type
but instead of an extended XY type.  
A geometrical explanation for the appearance of the n.n.n
and the next-next nearest neighbor (n.n.n.n) term is illustrated
in Fig.~1(a). Indeed, these terms come from the second trace
proportional to $t^{2}$, which has a factor
${\tilde\Sigma}^{(0)}{\tilde\Sigma}^{(0)}$. Each self-energy 
${\tilde\Sigma}^{(0)}$ picks up ${\bf \delta}=\pm{\hat x},\;\pm{\hat y}$,
and the second trace gives terms with resulting vectors
$\delta+\delta'=\delta_{2}$ or $\delta_{3}$,
where $\delta_{2}=\pm{\hat x}\pm{\hat y}$ and
$\delta_{3}\pm2{\hat x},\;\pm2{\hat y}$. 
This geometrical picture
also works when we include the n.n.n hopping $(t')$ in the electron
dispersion curves.

The physics of $\H_{\theta}$ depends on the relative magnitudes of
coefficients $J_{1}$, $J_{2}$, and $J_{3}$ as well as their relative signs.
For example, if $\beta$ is negligible, and $\alpha$ and $\gamma$ are both
positive so that $J_{1}>2J_{2}$,
$\H_{\theta}$ describe a non-frustrated XY model, and its
critical behavior can be understood\cite{simon} 
in terms of the usual XY Hamiltonian
with an effective coupling constant $J_{eff}=(J_{1}-2J_{2})$.
However, this does not mean that in this case
$\H_{\theta}$ is equivalent to
$-J_{eff}\sum_{<ij>}{\bf S}_{i}\cdot{\bf S}_{j}$
because
the local behaviors of these two Hamiltonians are different.
In general, however, as long as $J_{1}$ is dominant, the large length scale
behavior is of the usual XY type.
To calculate $J_{1}$, $J_{2}$, and $J_{3}$,
we need to know how $\Delta$ and $\mu$ change with increasing
temperature.
We choose the pairing interaction $U=1.4t$ and the filling factor
$n=0.9$ and self-consistently solve the equations
for gap and filling factor
to determine $\Delta(T)$ and $\mu(T)$.
What we obtain in the numerical calculation is that
i) $J_{3}$ is negligible from $T=0$ to $T=T_{MF}$, and ii)
the inequality
$J_{1}>2J_{2}$ holds almost all the way
up to $T=T_{MF}$. Near $T_{MF}$, $J_{1}$ can be
a little less than $2J_{2}$; however, as we mentioned before
the local effective Hamiltonian for phase fluctuations
may not be valid in this temperature regime, where effects of
thermally excited quasiparticles cannot be neglected.
Consequently, for the interesting temperature range
$J_{eff}=J_{1}-2J_{2}$; namely, $J_{eff}=2(\alpha+\beta)$.
Interestingly, $J_{eff}$ does not depend on the coefficients of 
the n.n.n and the n.n.n.n interaction terms.
However, this is not the case if we include the n.n.n hopping.

When we include the n.n.n hopping $(t')$, 
the Green function $G_{0}(i,j)$ and
the self energy change as follows:
$G^{-1}_{0}(i,j)\rightarrow G^{-1}_{0}(i,j)+{\hat\tau}_{3}
t'\sum_{\delta_{2}}\delta_{j,i+{\bf\delta}_{2}}$, and
$\Sigma^{(0)}(i,j)\rightarrow \Sigma^{(0)}(i,j)+
it'\sum_{\delta_{2}}\delta_{j,i+{\bf\delta}_{2}}
\sin(\theta_{i,j}/2)$,
and a similar expression for $\Sigma^{(3)}(i,j)$ holds.
In the same way that 
the $t^{2}$ term gives interactions of the form
${\bf S}_{i}\cdot{\bf S}_{i+\delta_{2}}$ and 
${\bf S}_{i}\cdot{\bf S}_{i+\delta_{3}}$, the $tt'$ term induces the form
${\bf S}_{i}\cdot{\bf S}_{i+\delta_{4}}$ because 
$\delta+\delta_{2}=\delta_{4}$, and the $t'^{2}$ term gives rise to
${\bf S}_{i}\cdot{\bf S}_{i+\delta_{3}}$ and
${\bf S}_{i}\cdot{\bf S}_{i+\delta_{5}}$
because $\delta_{2}+\delta'_{2}=\delta_{3}$ or $\delta_{5}$, where
$\delta_{4}=\pm{\hat x}\pm2{\hat y},\;\pm2{\hat x}\pm{\hat y}$, and
$\delta_{5}=\pm2{\hat x}\pm2{\hat y}$. 
Geometrical descriptions for these terms are also presented in Fig.~1(b) and
Fig.~1(c).
Following the same manipulation
as before, we obtain the effective Hamiltonian including
the n.n.n hopping contribution:
\begin{small}
\bea
\H_{\theta}=&&-{\cal J}_{1}\sum_{<ij>}{\bf S}_{i}\cdot{\bf S}_{j}
+{\cal J}_{2}\sum_{\ll ij\gg}{\bf S}_{i}\cdot{\bf S}_{j}
+{\cal J}_{3}\sum_{<ij>_{3}}{\bf S}_{i}\cdot{\bf S}_{j}
\nonumber\\
&&+{\cal J}_{4}\sum_{<ij>_{4}}{\bf S}_{i}\cdot{\bf S}_{j}
+{\cal J}_{5}\sum_{<ij>_{5}}{\bf S}_{i}\cdot{\bf S}_{j}\;,
\label{extH}
\eea
\end{small}
where $\sum_{<ij>_{4(5)}}$ means $\sum_{i,\delta_{4(5)}}$.
It is obvious that the extended features of $\H_{\theta}$ are
robust in band structure parameters. 
Moreover, contrary to what we found before, now 
${\cal J}_{2}$ is finite even at $T=0$ because
the n.n.n hopping $(t')$ contributes to the first trace as well as
to the second trace\cite{later}. In order to calculate
${\cal J}_{i}\;(i=1,2,3,4)$ we choose $t'=-0.2t$ with the same values 
of parameters as before $(U=1.4t$ and $n=0.9)$.
Since our numerical calculation indicates that
${\cal J}_{1}$ and ${\cal J}_{2}$ are dominant and 
${\cal J}_{1}>2{\cal J}_{2}$ in most of the temperature range, we neglect 
${\cal J}_{3}$, ${\cal J}_{4}$, and ${\cal J}_{5}$. 
In this case as we mentioned earlier, the critical behavior
is still of the usual XY type so that we introduce an effective coupling
${\cal J}_{eff}={\cal J}_{1}-2{\cal J}_{2}$. In Fig.~2,
we plot ${\pi\over2}{\cal J}_{eff}$ vs $T$ scaled by $T_{MF}$. 
Since ${\cal J}_{eff}<0$ near $T_{MF}$, 
one might think this implies a frustrated XY case;
however, we again point out
that $\Delta(T)$ is comparable to $T$ in this region
and, therefore, we estimate
the validity of the local effective Hamiltonian to be $0<T<T_{\theta}$,
where $T_{\theta}$ is a temperature such that ${\cal J}_{eff}>0$.
We also considered a case of a stronger interaction
$(U=2t)$ and a lower filling factor $(n=0.4)$
with the same $t'$.
The behavior of ${\cal J}_{eff}$ is similar to the one in Fig.~2.
We emphasize that ${\cal J}_{eff}$ approaches zero earlier than
does the gap $\Delta(T)$. Note that in the continuum limit\cite{loktev}
the coupling constant vanishes at $T_{MF}$. 

In order to find out the BKT transition temperature $(T_{BKT})$, one has to
solve self-consistently $T_{BKT}={\pi\over2}{\cal J}_{eff}(T_{BKT})$ with
equations for the gap and the filling factor. Since $\Delta(T)$ and $\mu(T)$
have been self-consistently obtained for given parameters, 
the solution of $T={\pi\over2}{\cal J}_{eff}(T)$ 
(the crossing point between the dashed line and the solid curve)
gives $T_{BKT}$, which is
indicated by an arrow in Fig.~2. Suppose one use the usual XY Hamiltonian
including only ${\cal J}_{1}$ term in Eq.~\ref{extH}, as seen (dotted curve)
in Fig.~2, $T_{BKT}$ is larger than $T_{MF}$, which is unreasonable.
Corrections to the usual XY model are essential in our approach.

In summary, 
An effective Hamiltonian for phase fluctuations has been derived
starting from an attractive Hubbard model on a lattice.
Unlike the common assumption, we found that the effective Hamiltonian is not
of the usual XY type but is extended XY. This extended feature
is robust and has important consequences. It is reinforced 
when the next nearest neighbor hopping is considered.
The critical behavior is still BKT-like but with an
effective coupling which depends significantly 
on the temperature.

\begin{acknowledgments}
This work was supported in part by the National Sciences and the Engineering
Research Council of Canada (NSERC) and by the Canadian Institute for Advanced
Research (CIAR).
\end{acknowledgments}

\bibliographystyle{prl}

\begin{thebibliography}{1}

\bibitem{timusk} T. Timusk and B. Statt, Rep. Prog. Phys. {\bf 62},
61 (1999) and references therein.

\bibitem{ding} H. Ding \etal, Nature (London) {\bf 382}, 51 (1996);
A. G. Loeser \etal, Science {\bf 273}, 325 (1996).

\bibitem{emery} V. J. Emery and S. A. Kivelson, Nature (London) {\bf 374},
434 (1995).

\bibitem{levin} Q. J. Chen, I. Kosztin, B. Jank${\acute {\mbox o}}$, 
and K. Levin,
Phys. Rev. Lett. {\bf 81}, 4708 (1998).

\bibitem{mermin} N. D. Mermin and H. Wegner, Phys. Rev. Lett. {\bf 17},
1133 (1966).

\bibitem{kt} V. L. Berezinskii, Sov. Phys. JEPT {\bf 32}, 493 (1971);
J. M. Kosterlitz and D. J. Thouless, J. Phys. C {\bf 6},
1181 (1973).

\bibitem{negele} See, for example, J. W. Negele and H. Orland,
{\it Quantum Many-Particle Systems} (Addison-Wesley, 1988).

\bibitem{loktev} V. M. Loktev, R. M. Quick, and S. Sharapov,
Phys. Rep. {\bf 349}, 1 (2001) and references therein.

\bibitem{abrikosov} See, for example, A. A. Abrikosov, L. P. Gorkov, and
I. E. Dzyaloshinski, {\it Methods of Quantum Field Theory in Statistical
Physics} (Dover, New York, 1975).

\bibitem{aitchison} I. J. R. Aitchison, G. Metikas, and D. Lee,
Phys. Rev. B {\bf 62}, 6638 (2000).

\bibitem{sharapov} S. G. Sharapov, H. Beck, and V. M. Loktev,
Phys. Rev. B {\bf 64}, 134519 (2001). 

\bibitem{simon} P. Simon, J. Phys. A {\bf 30}, 2653 (1997).

\bibitem{later} We will present detailed expressions for ${\cal J}_{i}\;
(i=1,2,3,4)$ in a later publication.

\end{thebibliography}


\begin{figure}[tbp]
\includegraphics[width=\linewidth]{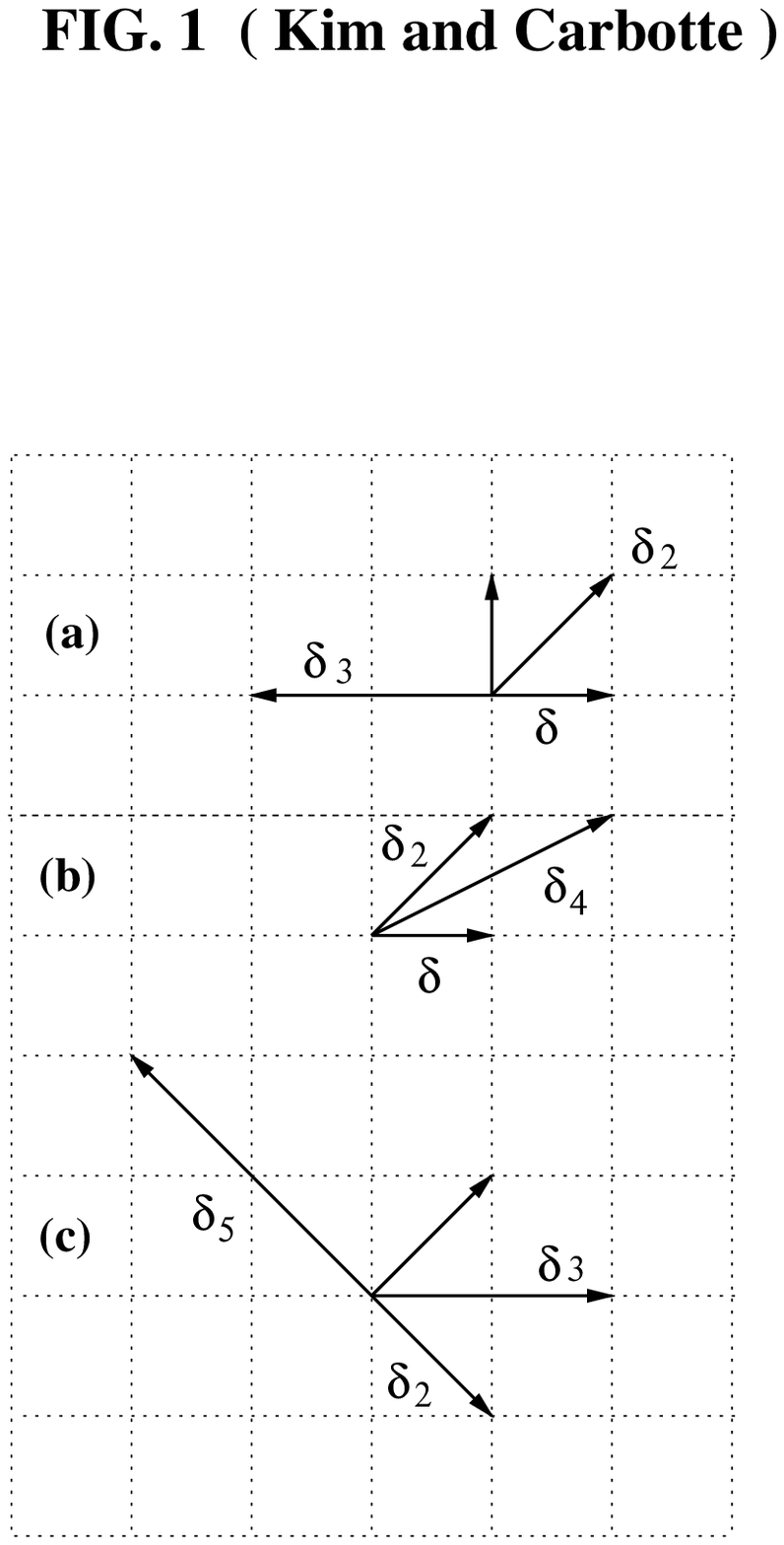}
\caption{
Geometrical illustrations of induced interactions
for the next-nearest and the further-neightbor spin pairs.
}
\end{figure}
\begin{figure}[tbp]
\includegraphics[width=\linewidth]{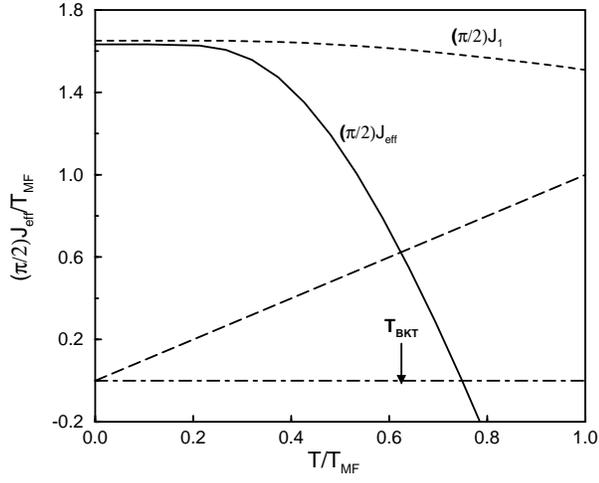}
\caption{
${\pi\over2}{\cal J}_{eff}$ (solid curve) as a function of temparature $(T)$
scaled by $T_{MF}$. The BKT temperature $T_{BKT}$ is indicated by an arrow.
If one use the usual XY Hamiltonian including only ${\cal J}_{1}$ term,
$T_{BKT}$ is larger than $T_{MF}$ (See the dotted curve).
}
\end{figure}

\end{document}